\documentclass[fleqn,10pt]{wlscirep}
\usepackage{lmodern}
\usepackage{graphicx}
\usepackage{amsmath}
\usepackage{amssymb}
\usepackage{bm}
\let\vec\bm

\title{Mass of Abrikosov vortex in high-temperature superconductor YBa$_2$Cu$_3$O$_{7-\delta}$}

\author[1,*]{Roman Tesa\v{r}}
\author[1]{Michal \v{S}indler}
\author[1]{Christelle Kadlec}
\author[2]{Pavel Lipavsk\'{y}}
\author[2]{Ladislav Skrbek}
\author[1]{Jan Kol\'{a}\v{c}ek}

\affil[1]{Institute of Physics, Czech Academy of Sciences, Na Slovance 2, 182 21 Prague 8, Czech Republic}
\affil[2]{Faculty of Mathematics and Physics, Charles University, Ke Karlovu 3, 121 16 Prague 2, Czech Republic}
\affil[*]{tesar@fzu.cz}


\begin{abstract}
Mass of Abrikosov vortices defied experimental observation for more than four 
decades. 
We demonstrate a method of its detection in high-temperature superconductors. 
Similarly to electrons, fluxons circulate in the direction given by the 
magnetic field, causing circular dichroism.
We report the magneto-transmittance of a nearly optimally doped thin 
YBa$_2$Cu$_3$O$_{7-\delta}$ film, measured using circularly polarized 
submillimeter waves. 
The circular dichroism emerges in the superconducting state and increases 
with dropping temperature. 
Our results confirm the dominant role of quasiparticle states in the vortex 
core and yield the diagonal fluxon mass of $2.2 \times 10^8$ electron masses 
per centimeter at 45~K  and zero-frequency limit and even larger off-diagonal 
mass of $4.9 \times 10^8 m_e$/cm. 
\end{abstract}

\begin{document}

\flushbottom
\maketitle
\thispagestyle{empty}


The limitations of today's digital technology will likely be overcome by 
ultra-fast superconductive electronics \cite{Anders10}, quantum circuits 
\cite{King18}, and quantum information processing. 
Such advanced electronic devices can operate within the framework of 
\emph{fluxonics} taking advantage of intrinsic properties of superconducting 
vortices. 
Abrikosov vortices, or fluxons, are ideal information carriers: they are 
topologically stable and keep a uniform nano-scale size given by quantum 
conditions. \cite{Golod15, Vlasko16}

Nevertheless, the fluxon mass is a particularly controversial issue. 
Its theoretical estimates are scattered over more than eight orders of 
magnitude.
Vortex mass in superconductors has been debated for over 50 
years since its theoretical prediction by Suhl \cite{Suhl65}, $4000~m_e/$cm 
for Nb, where $m_e$ is the electron mass. 
Multiple theories emerged in 1991, stimulated by the discovery of high-$T_c$ 
superconductors; for instance, the vortex mass in YBaCuO was predicted
\cite{CoffeyHao91} to reach $10^8\,m_e/$cm. 
Rather different trends follow from the theory of Han \emph{et al.} 
\cite{Han05}, which gives  $10^{13}\,m_e$/cm in Nb 
at 5~K, while for YBaCuO at the same temperature, it yields 
$3\times 10^9\,m_e$/cm.
Additionally, the vortex mass can be increased by backflow effects
\cite{Sonin13} or by the strain field \cite{Simanek91,Coffey94b} 
with estimates for YBaCuO from dominant $10^{10}$ to negligible $10^4\,m_e/$cm.
We also note that theories for the mass of quantized vortices in closely 
related superfluids offer similarly conflicting results 
\cite{Toikka17,Simula18} that span the full range from the infinite mass 
\cite{Popov73,Duan94a} to the negligible one \cite{Baym83}.

In contrast to the plethora of theoretical predictions of the effective 
vortex mass in type-II superconductors, we are aware of only two relevant 
experiments, both supporting rather large values. 
Fil \emph{et al.} \cite{Fil07} measured the acousto-electric effect in 
YB$_6$ and deduced a vortex mass of $10^{10}\,m_e/$cm.
Golubchik \emph{et al.} \cite{Golubchik12} observed the movement 
of individual vortices in a superconducting Nb film near its critical
temperature by combining high-resolution magneto-optical imaging with 
ultrafast heating and cooling. 
A nonzero mass is indispensable to explain their data; they estimated 
the fluxon mass to be in interval $(0.3 - 6) \times 10^8~m_e$/cm.

The extremely large spread in theoretically predicted values and scarce 
experimental data available motivated us to independently determine the 
mass of a fluxon. 
Here, we present relevant experiments leading to the evaluation of the vortex 
mass in nearly optimally doped YBa$_2$Cu$_3$O$_{7-\delta}$.
Our approach is inspired by the cyclotron resonance measurement, a method 
commonly employed to obtain the effective mass of charge carriers in 
semiconductors. 
Exposed to an external magnetic field, charged particles move in circular 
orbits and, under certain conditions, resonantly absorb monochromatic 
radiation.
In the case of fluxons, however, the cyclotron motion is induced by an 
interaction with circularly polarized far-infrared laser light. 
This interaction depends on the sense of circular polarization and manifests 
itself as a differential transmittance for clockwise and anti-clockwise 
polarized waves, known as magnetic circular dichroism.
We show that the theory of Kopnin and coworkers reproduces our experimental 
data \textit{without any fitting parameter}. 
Thus, our results support the theory of the fluxon mass developed by Kopnin 
and Vinokur \cite{KV98} and that of the Magnus force reduction by the factor 
of Kopnin and Kravtsov \cite{KK76}.

\section*{Far-infrared measurements of the fluxon mass}

\subsection*{Magneto-transmittance of circularly polarized light}

\begin{figure}[h]
\centering
\includegraphics[width=\linewidth]{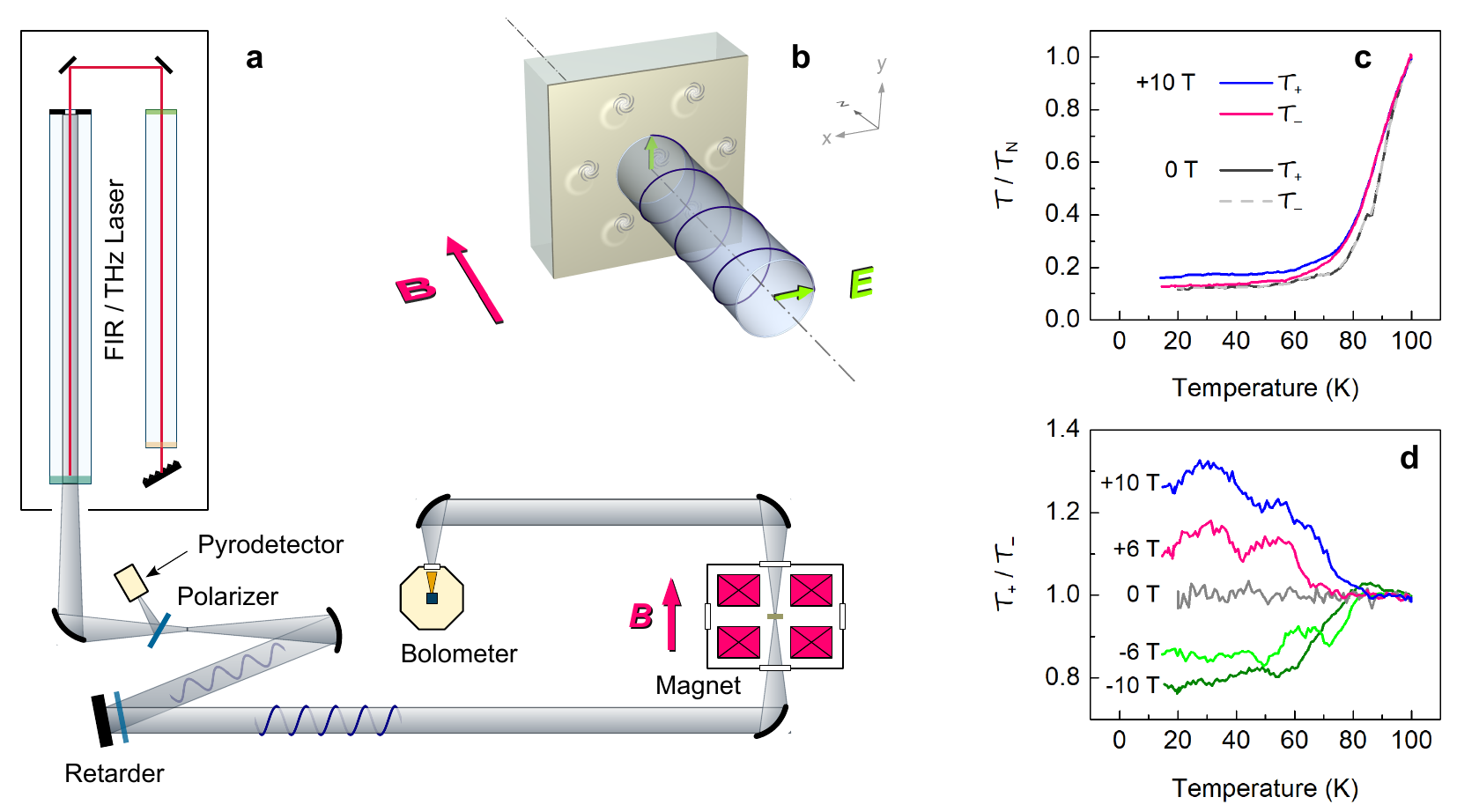}
\caption{Far-infrared measurement of fluxon mass.
({\bf a})~Sketch of the magneto-optical setup.
The continuous laser beam is split by a linear wire-grid polarizer;
the reflected part is monitored using a pyroelectric detector to keep trace 
of unavoidable power fluctuations, whereas the transmitted part proceeds 
toward the sample.
The retarder converts the light from linear to circular polarization. 
The propagation of the circularly polarized beam and the magnetic field 
are perpendicular to the film surface, as detailed in panel b. 
({\bf b})~Vortices in the film (gray circles) control the transmittance 
of the sample via the following mechanism: The electric field of the laser 
light drives the supercurrent. 
The Magnus force accelerates vortices in the direction perpendicular to the 
supercurrent; in reaction, the vortex motion affects the supercurrent and, 
thus, the transmittance. 
In the sketch, the electric field in the sample, as well as the vortices, 
rotate clockwise. 
If the light frequency is close to the cyclotron frequency of vortices, 
the motion of vortices is resonantly enhanced, leading to the observed 
dichroism. 
The extent of the cyclotron motion is strongly exaggerated; in fact, the 
fluxon circulates on a radius of less than $10^{-12}$ m at the strongest 
laser line.
({\bf c})~Transmittance of the YBa$_2$Cu$_3$O$_{7-\delta}$ superconducting 
sample, normalized to the normal-state transmittance ${\cal T}_N$ at 100~K 
and plotted for two circular polarizations versus temperature. 
The dichroism is clearly visible below 70~K in a magnetic field of 10~T;  
in a zero field, no dichroism appears. 
({\bf d})~Transmittance ratio ${\cal T}_+/{\cal T}_-$ measured in several 
applied magnetic fields plotted versus temperature.
The data were obtained using a 312 $\mu$m laser line ($6.1\times 10^{12}$ rad/s). 
Above the critical temperature, the dichroism is absent, showing that the 
normal-state Hall component is negligible.}
\label{fig:setup}
\end{figure}

We designed and developed a unique far-infrared (FIR) transmission 
experiment capable of probing the circular dichroism 
(see Fig.~\ref{fig:setup}). 
The breakthrough that has eventually allowed us to conduct this research is 
a custom-made retarder \cite{Tesar18} inserted in the optical path near the
FIR-laser output aperture; it delays the horizontal polarization relative to 
the vertical one by an adjustable phase difference.
A mutual phase shift of $\pm \pi/2$ converts the THz beam of equal vertical 
and horizontal polarization components into the circularly polarized state. 
Since the phase delay introduced by the retarder is inversely proportional
to the wavelength of the incoming light, each laser line requires a separate
adjustment.

We measured the transmittance of the sample for several laser lines at 
wavelengths 119, 163, 312, 419, and 433 $\mu$m \cite{Tesar10}, corresponding 
to the terahertz circular frequencies $\omega$ of 15.9, 11.6, 6.05, 4.50, 
and 4.35 $\times 10^{12}$ rad/s, respectively. 
Our setup shown in Fig.~\ref{fig:setup}a enables fast flips between clockwise 
($+$) and anti-clockwise ($-$) circular polarizations, consequently, we  
probed the transmittance ${\cal T}_+$ and ${\cal T}_-$ of both polarizations 
under identical conditions.
The transmittance, i.e., the fraction of laser energy transmitted through 
the sample, is evaluated as the bolometer-to-pyrodetector signal ratio, 
effectively eliminating any possible time instability in the laser power.
Identical profiles of both signals confirm that the transmittance is measured 
in the linear regime.

Our experimental protocol is as follows: in experimental runs, we apply a 
magnetic field at a temperature well above $T_c$. 
As the temperature drops, vortices freeze into a regular Abrikosov lattice. 
The applied magnetic field $B$ is kept constant since any change in $B$ would 
result in inhomogeneous patterns of the vortex density. 
The temperature $T$ is swept down and up at a steady sweep rate of 2.5 K/min, 
the instantaneous temperature of the sample being recorded together with the 
transmittance.
Care is taken that the down and up sweeps do not show any hysteresis.

Fig.~\ref{fig:setup}c displays a typical temperature behavior of the 
transmittance observed in the external magnetic field of 10~T using a 
312 $\mu$m laser line. 
The results obtained with different laser lines are similar.
As expected, the transmittance measured in zero field does not show any 
dichroism, in absence of Abrikosov vortices to induce an asymmetry.
In nonzero fields, however, the low-temperature circular dichroism is 
clearly observed and can be attributed to the formation of vortices 
threading the sample. 
The effect is enhanced in higher applied magnetic fields, thanks to the 
growing areal density of Abrikosov vortices (see Fig.~\ref{fig:setup}d). 
We focus on the low-temperature region, $T<50$~K, where the thermal 
quasiparticles play a negligible role and the response of the system 
is fully governed by the motion of fluxons.

\subsection*{Sample specification by time-domain terahertz spectroscopy }

\begin{figure}[h]
\centering
\includegraphics[width=\linewidth]{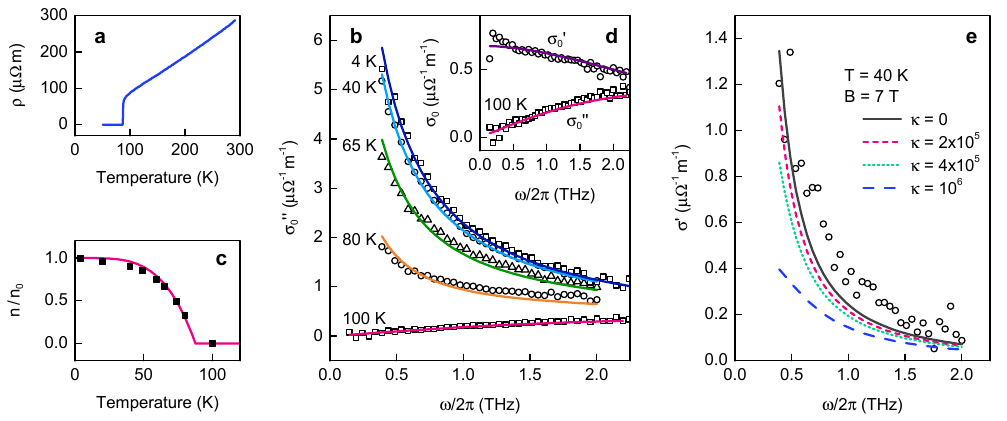}
\caption{Sample specification. Panels a--d refer to zero magnetic field.
(\textbf{a}) The dc resistivity as a function of 
temperature provides the critical temperature $T_c=87.6$~K. 
Extrapolation of the normal-state resistivity below $T_c$ gives negligible 
residual resistivity.
(\textbf{b}) The imaginary part of the conductivity as a function of frequency 
for several temperatures. 
The two-fluid model (lines) reproduces the data (symbols) for the 
superconducting fraction $f_s=1-T^4/T_c^4$ (line) compared in (\textbf{c}) 
with fitted values. 
(\textbf{d}) The real (circles) and imaginary (squares) conductivities at 
100 K provide the relaxation time $\tau_N=47$~fs. 
(\textbf{e}) The measured real part of conductivity for linearly polarized 
light at magnetic field 7~T (circles) compared with fits based on the vortex 
dynamics, $\sigma=\tfrac12(\sigma_++\sigma_-)$, supports small pinning with 
$\kappa=2\times 10^5$\,N/m$^2$.}
\label{fig:TDS}
\end{figure}

We chose the most common high-$T_c$ material YBa$_2$Cu$_3$O$_{7-\delta}$
in the form of a thin film with a thickness of $L = 107$ nm, and with CuO$_2$
planes parallel to the surface. 
The sample was prepared at National Chiao Tung University (Taiwan) using a 
pulsed laser deposition method from a stoichiometric target on a lanthanum
aluminate substrate oriented in the (100) plane.
The substrate dimensions are $10 \times 10 \times 0.5$ mm$^3$.
Several measurements were performed to establish sample properties.
Figure \ref{fig:TDS} summarizes some relevant results.

The critical temperature of the film, $T_c = 87.6$~K, was determined from 
a dedicated measurement of dc resistivity $\rho$ (Fig.~\ref{fig:TDS}a). 
For our slightly underdoped sample, this $T_c$ corresponds to the hole density 
\cite{Liang06} $n_0=1.68\times 10^{27}$/m$^3$. 
The linear slope of $\rho$ extrapolated to zero temperature shows negligible 
residual resistivity, so that the relaxation time is inversely proportional 
to temperature.

Additional film properties were established in a separate experiment using 
standard time-domain THz spectroscopy.  
Broadband linearly polarized THz pulses (0.3–-2.5 THz) were generated by 
exciting an interdigitated LT-GaAs emitter with a Ti:sapphire femtosecond 
laser beam at 800 nm~\cite{thz_setup2020}. 
We measured the complex conductivity $\sigma=\sigma'+i\sigma''$ for 
frequencies $\omega/2\pi$ in the range 0.5--2 THz at temperatures from 4 to 100~K 
(Figs.~\ref{fig:TDS}b,d,e). 
At the zero magnetic field and below $T_c$, the two-fluid model fit confirms the 
dominant London contribution, $\sigma_0\approx in e^2/(m\omega)$, with 
a temperature-dependent condensate density $n = n_0\,(1 - T^4/T_c^4)$,
the hole mass $m=3.3~m_e$, and the elementary charge $e$. 
Comparing the conductivities at temperatures of 4~K and $T_N=100$~K in 
Fig.~\ref{fig:TDS}d, we found the relaxation time $\tau_N=5\times 10^{-14}$~s.
Conductivities at all temperatures from 4 to 100 K are consistent with
$\tau = \tau_NT_N/T$.

Our sample is moderately clean. 
Its purity is given by the lifetime measured on the energy scale\cite{KV98}
as $k_{\rm B}^2T_c^2\tau/\hbar E_{\rm F}$, where $k_{\rm B}$ is the Boltzmann 
constant and $\hbar$ is the reduced Planck constant. 
The Fermi energy $E_{\rm F} = \hbar^2 k_{\rm F}^2/2m$ depends on the hole 
doping. 
For the hole density of our sample, the Fermi surface is cylindrical rather 
than spherical, and the Fermi momentum follows from the 2D density of holes 
in the CuO$_2$ plane $n_{\rm 2D} = n_0c/2 = k_{\rm F}^2/(2\pi)$, where 
$c = 11.68$~\AA~ is the YBaCuO lattice parameter in the $z$-direction. 
The resulting Fermi energy $E_{\rm F} = 71$~meV yields the value of
$k_{\rm B}^2T_c^2\tau/\hbar E_{\rm F} \sim 0.1$, corresponding to the 
moderately clean sample.

According to Kopnin and Vinokur \cite{KV98}, the moderately
clean d-wave superconductor behaves as the s-wave one. In the absence of 
reliable formulas for angular frequency $\omega_0$ of quasiparticles
bounded in the vortex core in the d-wave superconductors, 
we used $\hbar\omega_0=(\Delta_0/k_{\rm F}\xi_0)(1-T^2/T_c^2)$ for the 
conventional superconductors\cite{Soninbook}. 
We deduced the coherence length $\xi_0$ from the upper critical field 
in the zero-temperature limit \cite{Welp89}, 
$B_{c2} = 122~{\rm T}=\Phi_0/2\pi\xi_0^2$, where $\Phi_0$ is the magnetic 
flux quantum. 
From the energy gap $2\Delta_0 = 4.3 k_{\rm B}T_c$, we obtained 
$\omega_0 = 4.4\times 10^{12}$~rad/s, a value close to 
$4.5\times 10^{12}$~rad/s of our 419~$\mu$m laser line.

To complete the sample characteristics, we assume pinning of vortices 
by layer imperfections, for example, the surface roughness. 
Figure \ref{fig:TDS}e shows the conductivity in the magnetic field of 7~T 
applied perpendicularly to the film\cite{Sindler14, Kadlecf16}. 
It was interpreted either with the theory specified below or with the model 
used by Parks \cite{Parks95}, both indicating that the vortex pinning is 
rather weak with the Labusch coefficient 
$\kappa\approx 2\times 10^5$N/m$^2$.

\subsection*{Theoretical prediction}

\begin{figure}[h]
\centering
\includegraphics{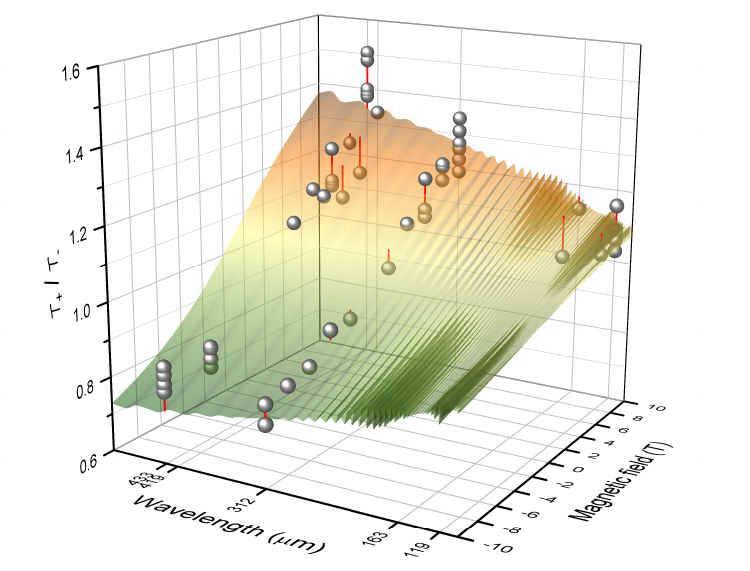}
\caption{Transmittance ratio ${\cal T}_+/{\cal T}_-$ for the laser lines 
119, 163, 312, 419, and 433 $\mu$m as a function of magnetic field. 
The theoretical prediction (colored surface) with no free fitting parameter 
is compared with experimental values (spheres) observed at a temperature 
of 45~K. The corrugation of the theoretical surface results from interference 
in the film/substrate structure. Its amplitude is smaller than the 
experimental error. The values observed under identical conditions but in
different runs hang on the same vertical line attached to the surface. 
These repeated measurements demonstrate a spread of the experimental data.
Despite the experimental uncertainty, the overall magnitude and dependence 
on magnetic field and wavelength are consistent with the Kopnin-Vinokur theory. }
\label{fig:3D}
\end{figure}

Our aim is to compare experimental values of ${{\cal T}_+}/{{\cal T}_-}$ 
with a theoretical model. Using the above sample parameters, we can evaluate 
the film conductivity $\sigma_\pm$ from which the transmittance ${\cal T}_\pm$ 
results. The theoretical prediction shown in Fig.~\ref{fig:3D} is based on 
the Yeh formalism \cite{Visnovsky} and covers interferences in the weakly 
birefringent substrate. For the purpose of discussion, we refer to an 
approximation ${{\cal T}_+}/{{\cal T}_-} = {|\sigma_-|^2}/{|\sigma_+|^2}$ 
which differs from the exact theory by less then 4\% as shown in the 
Supplementary Information \cite{suppl}.

Kopnin and Vinokur\cite{Kopnin01} provided the theoretical conductivities 
$\sigma_\pm$ derived under very general conditions. Our study allows for 
two simplifications. First, we focus on temperatures about 45 K, where 
we observe a large dichroic signal. At such low temperatures, extended 
quasiparticles are very dilute so that we can neglect their contribution 
to the electric current $\vec J$ as well as their effect on vortices. 
Second, the cyclotron frequency $\omega_c=eB/m$ is small on the scale of 
the quasiparticle lifetime, $\omega_c\tau\ll 1$ for all experimental 
magnetic fields, which simplifies dynamics of quasiparticles in the vortex
core. Under these conditions, the equation of motion for a fluxon of unitary 
length takes the form of Newton's law\cite{Kopnin01} 
\begin{equation}\label{forcebalance}
  \dot{\vec p} = \vec F 
    + \pi\hbar n [ (\vec J/(en)-\vec v) \times \vec z ]
    - \kappa\vec u 
\end{equation}
with the time derivative of momentum $\vec p$ and the force 
$\vec F$ from the interaction of the vortex core with the crystal lattice.
Unlike Kopnin and Vinokur, we include the pinning force with the Labusch 
parameter $\kappa$ and vortex displacement $\vec u$ related to its velocity 
as $\dot{\vec u} = \vec v$. 
The Magnus force, given by the vector product of the magnetic field direction 
$\vec z = \vec B/B$ and the vortex velocity related to the condensate current, 
covers a force by which the flowing condensate acts on the fluxon.

In Kopnin's model, the fluxon momentum $\vec p$ is a total momentum of 
quasiparticles in the vortex core which rotate about the vortex axis 
at an angular frequency $\omega_0$.
For our moderately clean sample, the fluxon momentum at the low-temperature 
limit depends on its velocity as \cite{KV98} 
$\vec p = \mu_\|\vec v-\mu_{\perp}[\vec v \times \vec z]$, where
\begin{equation} \label{mass}
  \mu_\|=\pi\hbar n \frac{\omega_0\tau^2}{(1 - i\omega\tau)^2 + \omega_0^2\tau^2} \,  
  ~~~~~~~ \mathrm{and} ~~~~~~~
  \mu_{\perp}=\pi\hbar n\tau \frac{1 - i\omega\tau}{(1 - i\omega\tau)^2 + \omega_0^2\tau^2} \, .
\end{equation}
The diagonal mass $\mu_\|$ is complex at finite frequencies, which reflects 
a delay between a change of the vortex velocity and a change of the total 
momentum of quasiparticles in its core. The off-diagonal mass $\mu_{\perp}$ 
describes a property common in anisotropic systems that the velocity of an 
excitation is not parallel with its momentum.

Quasiparticles disturbed by the vortex motion and action of the FIR light 
eventually lose their momentum in collisions with impurities and phonons. 
Via these collisions, the crystal lattice acts on the fluxon by force $\vec F$. 
With the collision integral approximated by a single relaxation time $\tau$, 
the force and the momentum are simply related by $\vec F=-\vec p/\tau$. 
The longitudinal force $-(\mu_\|/\tau)\vec v$ is a vortex friction. 
The transversal force $(\mu_\perp/\tau)[\vec v\times \vec z]$ reduces the 
Magnus force giving the Kopnin-Kravtsov force\cite{KK76} in the dc limit.

The electric field in the film\cite{Sonin96,Kopnin01,Lin12}
\begin{equation}\label{Josrel}
  \vec{E} = \frac1{\sigma_0} \vec{J} - [\vec v \times \vec B] 
\end{equation}
has a skin component $ \vec{J}/\sigma_0$ from the penetrating
incident light, and an electric field  $-[\vec v \times \vec B]$ 
generated by the vortex motion.
At low frequencies, the London conductivity $\sigma_0$ diverges and the electric 
field achieves the dc form of the Josephson relation. In the absence of 
the magnetic field, Eq.~\eqref{Josrel} reduces to the London formula. 
Alternatively, one can employ the Coffey-Clem model used by Parks 
{\em et al.}\cite{Parks95}, which is equivalent to Eq.~\eqref{Josrel}.

Conductivities $\sigma_\pm$ are diagonal elements of the conductivity tensor 
$\hat\sigma \vec E = \vec J$ in the helical basis. 
Eigen-vectors of circularly polarized light are constructed from basis 
vectors in the film plane, e.g., the electric field 
$\vec E = E_\pm \, {\rm e}^{-i \omega t}\vec e_\pm$,
where $\vec e_\pm=( {\vec x} \pm i s {\vec y})/\sqrt{2}$ are vectors of the 
helical basis and $s = {\rm sign}(\vec B\cdot\vec k)$ reflects the parallel 
or the anti-parallel orientation of the applied magnetic field $\vec B$ with 
respect to the wavevector $\vec k$ of incoming light.
The eigen-vectors satisfy $[\vec e_\pm \times {\vec z}] = \pm i s\vec e_\pm$, 
therefore, each of the vector equations \eqref{forcebalance} and \eqref{Josrel} 
splits into two independent scalar equations for $(+)$ and $(-)$ polarizations.
One can easily eliminate the velocity $\vec v$ and write the conductivities 
\begin{equation}\label{rhopm}
  \frac{1}{\sigma_\pm} = \frac{E_\pm}{J_\pm} = \frac{1}{\sigma_0}
  + \frac{B}{e n} \left( \frac{\omega_0\tau}{1 - i\omega\tau \mp i s {\omega_0\tau}}
                      + i\frac{\kappa}{\pi\hbar\omega n} \right)^{-1}
\end{equation}
needed for the theoretical prediction of the transmittance.

\section*{Discussion}

\subsection*{Experiment versus theoretical prediction}

With the full set of sample parameters established from the time-domain 
spectroscopy, the simplified Kopnin-Vinokur conductivity \eqref{rhopm}
furnishes us with the theoretical prediction of the circular dichroism. 
Figure~\ref{fig:3D} compares this prediction and the experimetally 
observed dichroism for several values of
the applied magnetic field and the THz-laser wavelength. 
The differences between theory and experiment are smaller than the 
experimental errors.

With increasing wavelength and field strength, the transmittance ratio 
${\cal T}_+/{\cal T}_-$ gradually deviates from unity. 
The observed trends can be understood in terms of Eq.~\eqref{Josrel}. 
The field dependence arises from the dominant London contribution $1/\sigma_0$ 
complemented by the Josephson-type resistivity, which is linear in $\vec B$. 
The variation with wavelength has a similar cause: for lower frequencies, the 
London resistivity $1/\sigma_0 \approx -i\omega m /( ne^2)$ is smaller so 
the Josephson part becomes dominant.

\subsection*{Vortex mass from experiment}

Figure~\ref{fig:3D} documents that the theory of Kopnin and Vinokur is 
relevant for the THz dynamics of vortices. Since the observed frequency 
dependence of magneto-transmission agrees with the theoretical one, even 
with \textit{no adjustable parameters}, the extrapolation of our results 
to low frequencies is justified. Based on this, we used their theory to 
find the vortex mass from our data.

While the sample parameters $n$, $\sigma_0$, and $\tau$ are sound, 
$\kappa$ and $\omega_0$ are less clear. The Labusch parameter $\kappa$ 
has a minor effect on the dichroism, therefore we kept the value 
$\kappa = 2 \times 10^5$ N/m$^2$.
The angular frequency $\omega_0$ was established by the least-square fit of 
${\cal T}_+/{\cal T}_-$ data. 
The best-fit value $\omega_0 = 4.3 \times 10^{12}$ rad/s was very 
close to $4.4\times 10^{12}$ rad/s estimated above. We believe that such 
close agreement of observed and estimated frequency is fortuitous.

At THz frequencies, where the circular dichroism was found, both diagonal 
and off-diagonal masses are complex. They become real in the low-frequency 
limit, as apparent from Eq. \eqref{mass}. 
Using the experimentally established values of 
$\omega_0 = 4.3\times 10^{12}$~rad/s and other sample parameters, we 
evaluated the zero-frequency components of the fluxon mass. 
In YBa$_2$Cu$_3$O$_{7-\delta}$ at 45~K, the diagonal mass $\mu_\|$ 
amounts to $2.2 \times 10^8~m_e$/cm, while the off-diagonal mass 
$\mu_\perp$ is more than twice larger, $4.9 \times 10^8~m_e$/cm.

In summary, we have developed a reliable experimental method to measure the 
circular dichroism of superconducting films threaded by Abrikosov vortices. 
To interpret our data in terms of vortex dynamics, we have established all 
the essential material parameters from independent time-domain THz spectroscopy 
and dc-resistivity measurements. 
The observed dichroism is in good agreement with the theory of 
Kopnin and Vinokur based on the circular motion of quasiparticles in the 
vortex core. Their angular frequency was experimentally determined and 
used to extrapolate the vortex mass from THz frequencies to low-frequency 
motion.

\bigskip

\section*{Acknowledgements}
We thank Chih-Wei Luo for preparing the YBaCuO sample and for providing
the temperature dependence of dc resistivity. 
The authors are grateful to E.~B.~Sonin for his kind help with the 
finite-frequency modification of the equation of vortex motion and to 
T. Simula and G. E. Volovik for valuable comments. 
We acknowledge the support of the Czech Science Foundation (project No. 21-11089S), 
P. L. that of INTER-EXCELLENCE (COST) LTC 18024, 
C. K. SOLID21-CZ.02.1.01/0.0/0.0/16 019/0000760 and 
L. S. thanks the Czech Science Foundation (project No. 20-00918S).

\section*{Author contributions statement}
J.K. designed the experiment;
R.T. developed the circular polarizer and built the experimental setup;
R.T. and M.\v{S}. performed the measurement of dichroism;
C.K. performed the time-domain THz measurements;
P.L., M.\v{S}., L.S., and J.K. analyzed the results.
All authors contributed to the discussions and production of the manuscript.

\section*{Additional information}
The authors declare no competing interests.


\begin{thebibliography}{10}

\bibitem{Anders10}
S. Anders \emph{et al.},
European roadmap on superconductive electronics -- status and perspectives,
Physica C \textbf{470}, 2079 (2010).

\bibitem{King18}
A. D. King \emph{et al.},
Observation of topological phenomena in a programmable lattice of 1,800 qubits,
Nature \textbf{560}, 456 (2018).

\bibitem{Golod15}
Golod, T., Iovan, A., and Krasnov, V. 
Single Abrikosov vortices as quantized information bits. 
Nat Commun \textbf{6}, 8628 (2015). https://doi.org/10.1038/ncomms9628

\bibitem{Vlasko16}
V. K. Vlasko-Vlasov, F. Colauto, T.  Benseman, D. Rosenmann, and W.-K. Kwok
Triode for Magnetic Flux Quanta. 
Sci Rep \textbf{6}, 36847 (2016). https://doi.org/10.1038/srep36847

\bibitem{Suhl65}
H. Suhl,
Inertial mass of a moving fluxoid,
Phys. Rev. Lett. \textbf{14}, 226 (1965).

\bibitem{CoffeyHao91}
M. W. Coffey and Z. Hao,
Dipolar electric field induced by a vortex moving in an anisotropic superconductor,
Phys. Rev. B \textbf{44}, 5230 (1991).

\bibitem{Han05}
J. H. Han, J. S. Kim, M. J. Kim, and P. Ao,
Effective vortex mass from microscopic theory,
Phys. Rev. B \textbf{71}, 125108 (2005).

\bibitem{Sonin13}
E. B. Sonin,
Transverse force on a vortex and vortex mass: Effects of free bulk and vortex-core bound quasiparticles,
Phys. Rev. B \textbf{87}, 134515 (2013).

\bibitem{Simanek91}
E. \v{S}im{\'a}nek,
Inertial mass of a fluxon in a deformable superconductor,
Phys. Lett. A \textbf{154}, 309 (1991).

\bibitem{Coffey94b}
M. W. Coffey,
Deformable superconductor model for the fluxon mass,
Phys. Rev. B \textbf{49}, 9774 (1994).

\bibitem{Toikka17}
L. A. Toikka and J. Brand,
Asymptotically solvable model for a solitonic vortex in a compressible superfluid,
New J. Phys. \textbf{19}, 023029 (2017).

\bibitem{Simula18}
T. Simula,
Vortex mass in a superfluid,
Phys. Rev. A \textbf{97}, 023609 (2018).

\bibitem{Popov73}
V. N. Popov,
Quantum vortexes and phase-transition in bose systems,
Zh. Eksp. Teor. Fiz. \textbf{64}, 672 (1973). [Sov. Phys. JETP \textbf{37}, 341 (1973)].

\bibitem{Duan94a}
J. M. Duan,
Mass of a vortex line in superfluid $^{4}\mathrm{He}$: Effects of gauge-symmetry breaking,
Phys. Rev. B \textbf{49}, 12381 (1994).

\bibitem{Baym83}
G. Baym and E. Chandler,
The hydrodynamics of rotating superfluids. I. zero-temperature, nondissipative theory,
J. Low Temp. Phys. \textbf{50}, 57 (1983).

\bibitem{Fil07}
V. D. Fil,  T. V. Ignatova, N. G. Burma, A. I. Petrishin, D. V. Fil, and N. Yu. Shitsevalova,
Mass of an Abrikosov vortex,
Low Temp. Phys. \textbf{33}, 1019 (2007).

\bibitem{Golubchik12}
D. Golubchik, E. Polturak, and G. Koren,
Mass of a vortex in a superconducting film measured via magneto-optical imaging plus ultrafast heating and cooling,
Phys. Rev. B \textbf{85}, 060504 (2012).

\bibitem{KV98}
N. B. Kopnin and V. M. Vinokur,
Dynamic vortex mass in clean Fermi superfluids and superconductors,
Phys. Rev. Lett. \textbf{81}, 3952 (1998).

\bibitem{KK76}
N. B. Kopnin and V. E. Kravtsov,
Conductivity and Hall effect of pure type-II superconductors at low temperatures,
Pis'ma Zh. Eksp. Teot. Fiz. \textbf{23}, 631 [JETP Lett. \textbf{23}, 578] (1976).

\bibitem{Tesar18}
R. Tesa\v{r},  M. \v{S}indler, J. Kol\'a\v{c}ek, and L. Skrbek,
Terahertz wire-grid circular polarizer tuned by lock-in detection method,
Rev. Sci. Instrum. \textbf{89}, 083114 (2018).

\bibitem{Tesar10}
R. Tesa\v{r}, Z. \v{S}im\v{s}a, J. Kol\'a\v{c}ek, M. \v{S}indler, L.~Skrbek, K. Il'in, and M. Siegel,
Terahertz transmission of NbN superconductor thin film,
Physica C \textbf{470}, 932 (2010).

\bibitem{Liang06}
R. Liang,  D. A. Bonn, and W. N. Hardy,
Evaluation of CuO$_2$ plane hole plane hole doping in YBa$_2$Cu$_3$O$_{6+x}$ single crystals,
Phys. Rev. B \textbf{73}, 180505(R) (2006).

\bibitem{thz_setup2020}
N. Blumenschein, C. Kadlec, O. Romanyuk, T. Paskova, J. F. Muth, and F. Kadlec, 
Dielectric and conducting properties of unintentionally and Sn-doped $\beta$-Ga$_2$O$_3$ studied by terahertz spectroscopy, 
J. Appl. Phys. \textbf{127}, 165702 (2020).

\bibitem{Soninbook}
E. B. Sonin, 
{\it Dynamics of Quantised Vortices in Superfluids}, 
Cambridge University Press (2016).

\bibitem{Welp89}
U. Welp,  W. K. Kwok, G. W. Crabtree, K. G. Vandervoort, and J. Z. Liu,
Magnetic measurements of the upper critical field of YBa$_2$Cu$_3$O$_{7-\delta}$ single crystals,
Phys. Rev. Lett. \textbf{62}, 1908 (1989).

\bibitem{Sindler14}
M. \v{S}indler, R. Tesa\v{r},  J. Kol\'a\v{c}ek, P. Szab\'{o}, P. Samuely, V. Ha\v{s}kov\'{a}, C. Kadlec, F. Kadlec, and P. Ku\v{z}el, 
Superdcond. Sci. Technol., \textbf{27}, 055009 (2014).

\bibitem{Kadlecf16}
F. Kadlec, C. Kadlec, J. V\'it, F. Borodavka, M. Kempa, J. Prokle\v{s}ka, J. Bur\v{s}\'ik, R. Uhreck\'y, S. Rols, Y. S. Chai, K. Zhai, Y. Sun, J. Drahokoupil, V. Goian, and S. Kamba, 
Electromagnon in the $Z$-type hexaferrite (Ba$_x$Sr$_{1-x}$)$_3$Co$_2$Fe$_{24}$O$_{41}$,
Phys. Rev. B, \textbf{94}, 024419 (2016).

\bibitem{Parks95}
B. Parks, S. Spielman, J. Orenstein, D. T. Nemeth, F. Ludwig, J. Clarke, P. Merchant, and D. J. Lew,
Phase-sensitive measurements of vortex dynamics in the terahertz domain,
Phys. Rev. Lett. \textbf{74}, 3265 (1995).

\bibitem{Visnovsky}
\v{S}. Vi\v{s}\v{n}ovsk\'y,
\emph{Optics in Magnetic Multilayers and Nanostructures} (CRC Press, 2006).

\bibitem{suppl}
Supplementary Information

\bibitem{Kopnin01}
N. B. Kopnin and V. M. Vinokur, 
Superconducting Vortices in ac Fields: Does the Kohn Theorem Work?,
Phys. Rev. Lett. \textbf{87}, 017003 (2001).

\bibitem{Sonin96}
E. B. Sonin, 
Interaction of ultrasound with vortices in type-II superconductors,
Phys. Rev. Lett. \textbf{76}, 2794 (1996).

\bibitem{Lin12}
P.-J. Lin and P. Lipavsk{\'y} and P. Matlock, 
Inertial Josephson relation for FIR frequencies,
Phys. Lett. A \textbf{376}, 883 (2012).

\bibitem{GR66}
J. I. Gittleman and B. Rosenblum,
Radio-Frequency Resistance in the Mixed State for Subcritical Currents,
Phys. Rev. Lett. \textbf{16}, 734 (1966).





\end{thebibliography}
\end{document}


\noindent\LARGE\bf
Supplementary Information
\medskip

\noindent\normalsize 
Mass of Abrikosov vortex in high-temperature superconductor YBa$_2$Cu$_3$O$_{7-\delta}$

\noindent\normalfont
Roman Tesa\v{r}, Michal \v{S}indler, Christelle Kadlec, Pavel Lipavsk\'{y}, Ladislav Skrbek, and Jan Kol\'{a}\v{c}ek

\section*{Calculation of transmittance}\label{App:Yeh}
\renewcommand{\theequation}{S\arabic{equation}}
\renewcommand{\thefigure}{S\arabic{figure}}
\setcounter{equation}{0}
\setcounter{figure}{0}

\begin{figure}[h]
\centering
\includegraphics{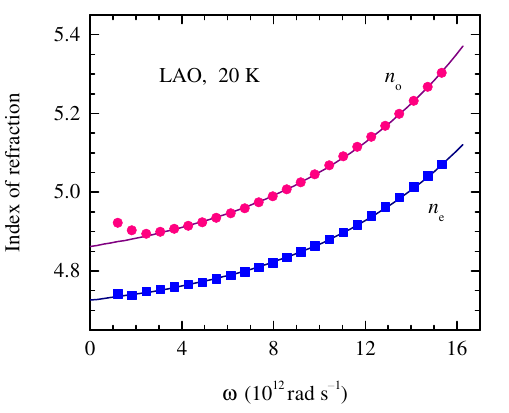}
\caption{Ordinary ($n_{\rm o}$) and extraordinary ($n_{\rm e}$)
refractive index of lanthanum aluminate (LAO) substrate measured at
a temperature of 20~K.
The frequency dependence is well described by exponential functions 
$n_{\rm o} = 4.789+0.073 \exp(0.127\,\omega)$ and 
$n_{\rm e} = 4.675+0.051 \exp(0.134\,\omega)$ with 
the angular frequency $\omega$ in $10^{12}$ rad/s units (solid lines).} 
\label{fig:LAO}
\end{figure}

Our sample comprised a thin YBaCuO film with a thickness of $L=107$~nm, 
deposited on a lanthanum aluminate (LAO) substrate. 
Anisotropic properties of the substrate were measured with standard 
time-domain THz spectroscopy. 
For convenience, the $x$-axis was chosen parallel to the linear polarization 
of the ordinary ray. 
At a low temperature of 20~K, we established the substrate thickness 
$D = 513.5~\mu$m and dispersion of ordinary and extraordinary refractive indices 
shown in Figure \ref{fig:LAO}.
Measurements at temperatures up to 100 K did not reveal any appreciable deviation 
from the low-temperature values.

When calculating the sample transmittance, it is necessary to combine 
two different approaches. 
Circular dichroism in YBaCuO is naturally described within a vector basis 
related to the circular polarization of the laser beam.  
On the other hand, the birefringence of the substrate is conveniently 
represented in a linear basis associated with the anisotropy axes of LAO. 
To match these two basis we employ Yeh's $4 \times 4$ matrix algebra 
\cite{Yeh79,Yeh80,Visnovsky,Sindler12}.

\begin{figure}[h]
\centering
\includegraphics[width=\textwidth]{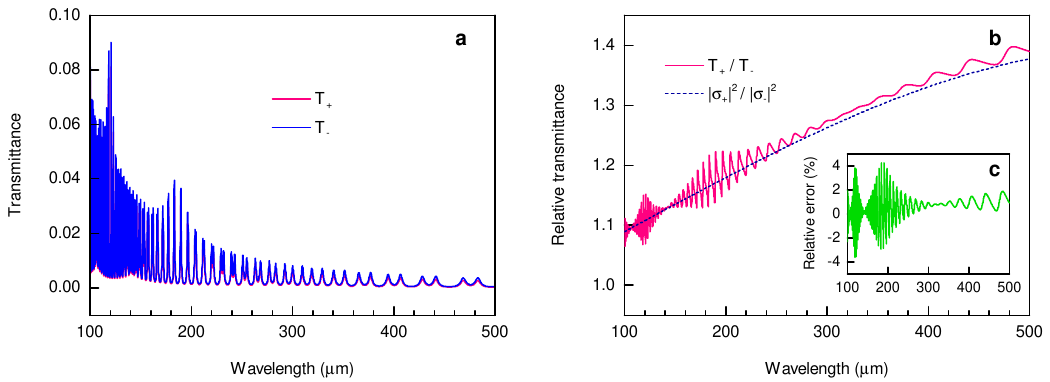}
\caption{Transmittance as a function of the laser wavelength for $B=10$~T 
and $T=45$~K. 
\textbf{(a)} Interference in the sample substrate leads to strong oscillation 
in the transmittance. Due to weak birefringence of the substrate, both 
circular polarizations follow similar but not identical pattern. 
\textbf{(b)} The interference almost cancels in the ratio of transmittances 
(red line). The free-film approximation (blue line) differs from the exact 
result by less than 4\%, as detailed in the inset \textbf{(c)}.} 
\label{fig:interfer}
\end{figure}

In each of the segments, vacuum$|$YBaCuO$|$LAO$|$vacuum, we write the 
resulting electric field $\vec E(z,t)=\vec E(z){\rm e}^{-i\omega t}$ as 
a sum of four partial waves\cite{Yeh79,Visnovsky}
\begin{equation}
  \label{eq:waves}
  {\vec E}(z) = \sum_{j=1}^4 E_j \, {\vec e}_j \,
  {\rm e}^{i k_j z}\, ,
\end{equation}
where $\vec{e}_j$ stand for eigen-polarization vectors and $k_j$ for 
wavevectors. 
Partial waves propagating forward are indexed by odd numbers $j = 1,\,3$ 
and waves propagating backward by even numbers $j = 2,\,4$.
In vacuum and YBaCuO, the eigen-polarization vectors $\vec{e}_j$ are given   
in the helical basis as 
$\vec e_{1,3}=\vec e_{2,4} = (\vec x\pm i\vec y)/\sqrt{2}$.
In LAO, we use the linear basis
$\vec e_{1, 2} = \vec x$ and 
$\vec e_{3, 4} = \vec y$.
Now we collect the amplitudes $E_j$ of partial waves into a four-component 
column vector and express the light propagation in a matrix form as 
\begin{equation}
  \label{eq:M-matrix}
  {\bf E}_{in} = {\bf M} \, {\bf E}_{tr} \, ,
\end{equation}
where the $4\times 4$ transfer matrix
\begin{equation}
  {\bf M} =
  {\bf D}_{\rm vac}^{-1}\,
  {\bf D}_{\rm film}^{\vphantom{X}}\, {\bf P}_{\rm film}^{\vphantom{X}}\, {\bf D}_{\rm film}^{-1}\,
  {\bf D}_{\rm subs}^{\vphantom{X}}\, {\bf P}_{\rm subs}^{\vphantom{X}}\, {\bf D}_{\rm subs}^{-1}\,
  {\bf D}_{\rm vac}\, 
\end{equation}
connects the incident electric field with the transmitted one via a sequence 
of dynamical and propagation matrices. 
The propagation matrices are diagonal. 
For the YBaCuO film, we have 
$({\bf P}_{\rm film})_{ij}=\delta_{ij} {\rm e}^{i k_j L}$ with 
$k_{1,2}=\pm n_+\omega /c$ and $k_{3,4}=\pm n_-\omega /c$, where 
$c$ is the vacuum speed of light and
\begin{equation}
  n_{\pm} = \sqrt{\frac{i \sigma_{\pm}}{\omega \varepsilon_0 }}
\end{equation}
is the complex refractive index related to the complex circular conductivity 
$\sigma_{\pm}$.
Similarly, for the LAO substrate, 
$({\bf P}_{\rm subs})_{ij}=\delta_{ij} {\rm e}^{i q_j D}$ with 
$q_{1,2}=\pm n_{\rm o}\omega/c$
and $q_{3,4}=\pm  n_{\rm e}\omega/c$.
Matching of electromagnetic waves at segment interfaces is covered by dynamical matrices
\begin{equation}
  {\bf D}_{\rm vac}  =  \frac{1}{\sqrt{2}} \left(
  \begin{array}{rrrr}
    1  &   1  &   1  &   1  \\
    1  &  -1  &   1  &  -1  \\
    i  &   i  &  -i  &  -i  \\
   -i  &   i  &   i  &  -i
  \end{array} \; \right ) , \qquad
  {\bf D}_{\rm film}  = \frac{1}{\sqrt{2}} \left(
  \begin{array}{cccc}
     1      &   1      &   1      &   1     \\
     n_+    &  -n_+    &   n_-    &  -n_-   \\
     i      &   i      &  -i      &  -i     \\
    -i n_+  &   i n_+  &   i n_-  &  -i n_-
  \end{array} \right )  ,
\end{equation}
\begin{equation} \nonumber
  {\bf D}_{\rm subs} = \left(
  \begin{array}{cccc}
    1         &   1         &   0         &  0     \\
    n_{\rm o} &  -n_{\rm o} &   0         &  0     \\
    0         &   0         &   1         &  1     \\
    0         &   0         &  -n_{\rm e} &  n_{\rm e}
  \end{array} \right ) .
\end{equation}
As already mentioned, we use the same helical basis for vacuum and YBaCuO, 
which leads to a similarity of ${\bf D}_{\rm vac}$ and ${\bf D}_{\rm film}$.
Since no light approaches the sample from its backside, 
$({\bf E}_{tr})_2 = 0$ and $({\bf E}_{tr})_4 = 0$, 
we obtain the transmittances of circularly polarized light in the form
\begin{eqnarray}
  \mathcal{T}_+&= &\left( |M_{33}|^2 + |M_{31}|^2 \right) S ,  \nonumber \\
  \mathcal{T}_-&= &\left( |M_{13}|^2 + |M_{11}|^2 \right) S , 
\end{eqnarray}
where $S = \left| M_{11}M_{33} - M_{13}M_{31} \right|^{-2}$ is a submatrix 
determinant. 
The superposition of partial waves in the substrate leads to strong 
oscillations with the wavelength (see Fig.~\ref{fig:interfer}a). 
These interference effects, however, are considerably reduced in the ratio 
of transmittances with opposite helicity 
\begin{equation}
  \frac{\mathcal{T}_+}{\mathcal{T}_-} = \frac{ |M_{33}|^2 + |M_{31}|^2 }
                                             { |M_{13}|^2 + |M_{11}|^2 },
\end{equation}
as documented in Fig.~\ref{fig:interfer}b. 

\bigskip
For purposes of discussion, we can use a free-film approximation which 
is obtained by sending the substrate thickness to zero, $D\to 0$. 
A further simplification follows from the expansion in the film thickness 
according to Hooper and Sambles\cite{Hooper08}, 
\begin{equation}\label{HS}
\frac{1}{\cal T} = 1
+ \frac{\sigma'}{\varepsilon_0 \omega}                \left( \frac{L\omega}{c} \right)
+ \frac{|\sigma|^2}{4 \varepsilon_0^2 \omega^2}       \left( \frac{L\omega}{c} \right)^2
+ \frac{\sigma' \sigma''}{3 \varepsilon_0^2 \omega^2} \left( \frac{L\omega}{c} \right)^3
+ \frac{1}{12}
  \left [  2  \frac{|\sigma|^2}{\varepsilon_0^2 \omega^2} +\frac{\sigma''}{\varepsilon_0 \omega} + \frac{\sigma'' |\sigma|^2}{(\varepsilon_0 \omega)^3} \right] 
  \left (\frac{L \omega}{c} \right)^4 ,
\end{equation}
adapted to our notation by substituting for permittivity 
$\varepsilon = 1 + \mathrm{i} \sigma/(\varepsilon_0 \omega)$. 
This expansion is nearly exact but still too complicated. 
Numerical calculations reveal that the third term is dominant in our case 
\begin{equation}\label{HSap}
{\cal T} \approx \frac{4 \varepsilon_0}{\mu_0 L^2 |\sigma|^2}.
\end{equation}
We used the approximation \eqref{HSap} in the main text to discuss the 
observed dichroism 
\begin{equation}
 \frac{{\cal T}_+}{{\cal T}_-} \,\approx\, \frac{|\sigma_-|^2}{|\sigma_+|^2} .  
\end{equation}
Its validity is demonstrated in Fig.~\ref{fig:interfer}b.

\bigskip

